\RequirePackage{ifpdf}
\ifpdf 
\documentclass[pdftex]{sigma}
\else
\documentclass{sigma}
\fi

\numberwithin{equation}{section}

\newcommand {\ch} {\mathcal{H}}

\newcommand{\dop}{\partial_t}
\newcommand{\zbar}{\overline{z}}

\begin{document}

\allowdisplaybreaks

\renewcommand{\thefootnote}{$\star$}

\renewcommand{\PaperNumber}{036}

\FirstPageHeading

\ShortArticleName{Models of Quadratic Algebras Generated by Superintegrable Systems in 2D}

\ArticleName{Models of Quadratic Algebras Generated\\ by Superintegrable Systems in 2D\footnote{This paper is a
contribution to the Special Issue ``Symmetry, Separation, Super-integrability and Special Functions~(S$^4$)''. The
full collection is available at
\href{http://www.emis.de/journals/SIGMA/S4.html}{http://www.emis.de/journals/SIGMA/S4.html}}}

\Author{Sarah POST}

\AuthorNameForHeading{S.~Post}

\Address{Centre de Recherches Math\'ematiques,
Universit\'e de Montr\'eal, \\C.P. 6128 succ. Centre-Ville, Montr\'eal (QC) H3C 3J7, Canada} 
\Email{\href{mailto:post@crm.umontreal}{post@crm.umontreal}}
\URLaddress{\url{http://crm.umontreal.ca/~post/}}

\ArticleDates{Received February 01, 2011, in f\/inal form March 24, 2011;  Published online April 05, 2011}

\Abstract{In this paper, we consider operator realizations of quadratic algebras generated by second-order superintegrable systems in 2D. At least one such realization is given for each set of St\"ackel equivalent systems for both degenerate and nondegenerate systems. In almost all cases, the models can be used to determine the quantization of energy and eigenvalues for integrals associated with separation of variables in the original system.}


\Classification{22E70; 81R05; 17B80}

\section{Introduction}

In his 1968 book {\em Lie theory and special functions}, W.~Miller Jr. used function space realizations of Lie algebras to establish a fundamental relationship between Lie groups and certain special functions including Bessel functions, hypergeometric functions and  conf\/luent hypergeometric functions~\cite{Miller68}. It was further shown how the algebra relations can be used to identify special function identities.

In this paper, we will apply these methods to the study of the representation theory for quadratic algebras generated by second-order superintegrable systems in 2D and their associated special functions. We would like to consider irreducible function space representations and so we restrict the Hamiltonian to a constant times the identity and construct dif\/ference or dif\/ferential operator realizations for the elements of the algebra. We call such operator realizations models.

A classical or quantum  Hamiltonian on an $m$ dimensional Riemannian manifold, with metric~$g^{jk}$,  given respectively by
\begin{gather*}
 \ch=\frac{1}{2} g^{jk}p_jp_k+V(x_1,x_2), \qquad H=\frac{-\hbar^2}{2 \sqrt{g}}\partial_{x_j}\sqrt{g}g^{jk}\partial_{x_k} + V(x_1,x_2)
 \end{gather*}
is called  superintegrable if it admits $2m-1$ integrals of motion. For classical systems, we require that the integrals be functionally independent and, for quantum systems, we require that that they be algebraically independent within a Jordan algebra generated by $x_j$, $\partial_{x_j}$ and the identity. If both of the integrals are polynomial in the momenta in the classical case and as dif\/ferential operators in the quantum case, we call the system $n$th-order superintegrable, where $n$ is the maximal order of a minimal generating set of integrals.

The study of superintegrability was pioneered by Smorodinsky, Winternitz and collaborators in the study of multiseparable systems on real Euclidean space \cite{FMSUW, MSVW1967}. They identif\/ied all four mulitseparable potentials on real Euclidean space, since named the Smorodinsky--Winternitz potentials. Later, W Miller Jr. with collaborators including E.~Kalnins, J.~Kress, and G.~Pogosyan published a series of papers which classif\/ied all second-order superintegrable systems in 2D (see e.g.~\cite{Eis, KKM20041, KKM20061} and references therein). In these paper, the authors proved that
classical and quantum second-order superintegrable systems are in one to one correspondence and that the potentials are invariant with respect to scaling so that it is possible to  normalize  the $-\hbar^2/2$ term to be~$1$.  Thus,  for the remainder of the paper we will only consider quantum systems, except to make slight observations where the classical systems dif\/fer, and chose this normalization.

It was further shown that any second-order superintegrable system in 2D can be related, via the St\"ackel transform, to a superintegrable system on a space of constant curvature and a~complete list was given of all second-order superintegrable systems  on 2D Euclidean space, $E_{2,\mathbb{C}}$, and on the two sphere, $S_{2, \mathbb{C}}$. Since we will not use the St\"ackel transform explicitly in this paper, we refer the reader to~\cite{KKM20042, Post20111} and references therein for a complete exposition. We only note here that the St\"ackel transform is a mapping between Hamiltonian systems, possibly on dif\/ferent manifolds, which preserves superintegrability and  the algebra structure of the integrals up to a~permutation of the parameters and the energy. Thus, we can classify superintegrable systems based on the structure of their symmetry algebra. Such classif\/ications have been worked out directly in~\cite{Dask2006, DaskTand2007} and via the St\"ackel transform~\cite{Kress2007, TD2008}.

Most importantly for this paper, it was proved that the algebra generated by the constants of the motion for a second-order superintegrable system closes to form a quadratic algebra. That is, suppose $H$ is second-order superintegrable with second-order integrals of the motion~$L_1$ and~$L_2 $. The integrals~$L_1$ and~$L_2$ will not commute and we denote their commutator by~$R$. The algebra to be considered is then the associative algebra generated by
\begin{gather*}
A=\left\{ L_1, L_2, H, R\equiv \left[ L_1,L_2\right] \right\}.
\end{gather*}
Such an algebra is called a quadratic algebra if the commutator of any two elements can be written as at most a quadratic polynomial in the generators. Further, since the four generators can not be independent in the Jordan algebra generated by the operators~$x_j$ and $\partial_{x_j}$ and the identity, there will be a polynomial relation between them.

The structure relations for our algebra are hence given by
\begin{gather} 
\label{a1}
 \left[L_1 , R\right]  =  P(L_1, L_2, H),\qquad
\left[L_2 , R\right]  = Q(L_1,L_2,H), \qquad
R^2  = S(L_1,L_2,H),
\end{gather}
where $P$ and $Q$ are at most quadratic polynomials and~$S$ is at most cubic.
We note that in the classical case, we will instead have a quadratic  Poisson algebra and the highest order terms of the structure relations will be the same as in  to the quantum case. If the system admits a~f\/irst-order integral~$X$, we call the system degenerate. In this case, the potential depends only on a single parameter and~$R$ can be expressed in terms of the basis $X$, $L_1$, $L_2$ and~$H$~\cite{KKMP1param}.

The study of the the algebras generated by superintegrable systems and in particular their representation theory has been a subject of recent study (see e.g. \cite{Dask2001, Dask2008, KMPost2010,  Marq2010, Quesne2007}). The main advantage of the method of function space realization is that it can be used to f\/ind the eigenvalues of operators other than the Hamiltonian and to compute inter-basis expansion coef\/f\/icients for the  wave functions of the Hamiltonian. Further, as in the case of Lie algebras, the representation theory for quadratic algebras, and polynomial algebras more generally, seems to be intimately connected with special functions and their identities.

This paper is divided up into four sections. In Section~\ref{section2}, we give a model which realizes the quadratic algebra associated to the singular isotropic oscillator, E1 or Smorodinsky--Winter\-nitz~i. A function space representation is given including the normalization and the weight function as well as the eigenfunctions and eigenvalues for each of the operators associated with separation of variables in the original system. We describe how the energy values for the system and eigenvalues of the   operators associated to separation of variables are quantized in a f\/inite dimensional representation. In Sections~\ref{section3} and~\ref{section4}, we give at least one model for the algebras associated with a representative of each St\"ackel equivalence class of nondegenerate and degenerate systems respectively. In these sections, the model is given and we identify possible f\/inite dimensional representations including those are associated with quantized values of the energy.  In Section~\ref{section5} there is a brief conclusion with possible future developments.

\section{Exposition of a model: the  singular isotropic oscillator}\label{section2}

In this section, we shall consider the quadratic algebra associated with the system E1/Smo\-ro\-dins\-ky--Winternitz~i and construct an irreducible function space representation for algebra. The Hamiltonian for this system, the singular isotropic oscillator, is
\[
 H= \partial_x^2+\partial_y^2 -\omega^2\big(x^2+y^2\big)+\frac{\frac14-a^2}{x^2}+\frac{\frac14-b^2}{y^2}.
 \]
The remaining generators of the symmetry algebra are,
\[L_1=\partial_x^2+\frac{1/4-a^2}{x^2}-\omega^2x^2, \qquad L_2=(x\partial_y-y\partial_x)^2+\frac{(1/4-a^2)y^2}{x^2}+\frac{(1/4-b^2)x^2}{y^2}.\]
The algebra relations are, recall $R\equiv \left[L_1, L_2 \right],$
\begin{gather}\label{E1alg1}
\left[R,L_1\right] =8L_1^2-8HL_1+16\omega^2L_2-8\omega^2,\\
\left[R,L_2\right] =8HL_2-8\lbrace L_1, L_2 \rbrace+\big(12-16a^2\big)H+\big(16a^2 +16b^2-24\big)L_1, \\
R^2 =8H\lbrace L_1,L_2\rbrace-\frac{8}{3}\lbrace L_1,L_1, L_2\rbrace +16\omega^2L_2^2+16\big(a^2-1\big)H^2+\left(16a^2+16b^2-\frac{200}3\right)L_1^2\nonumber\\
\phantom{R^2 =}{}  -\left(32a^2-\frac{200}3\right)HL_1-\frac{176\omega^2}{3}L_2-\frac{4\omega^2}{3}\big(48a^2b^2-48a^2-48b^2+29\big).\label{E1alg2}
\end{gather}

In order for the representation to be irreducible, the Hamiltonian will be restricted to a~constant and so, in the model, the action of the  Hamiltonian is represented by a constant $E$, i.e.~$H\equiv E$.
Now, suppose that $L_1$ is diagonalized by monomials, that is, it is of the form $L_1=l_1t\partial_t+l_0$ where $l_1$ and $l_0$ are some constants to be determined later. Suppose also that the action of $L_2$ on the basis of monomials is in the form of a three-term recurrence relation. From the algebra relations, it is straightforward to see that $L_2$ cannot be modeled by a f\/irst-order operator or else the relations would be a Lie algebra. Thus, we take the Ansatz of a second-order dif\/ferential operator with polynomial coef\/f\/icients.

The algebra relations (\ref{E1alg1})--(\ref{E1alg2}) are realized by the following operators, $H=E$ and
\begin{gather*}
L_1 =-4\omega t \partial_t+E-2\omega(1+b),\\
L_2 =\frac12t(8t+1)^2\partial_t^2-\left(\frac{16(E-2\omega b-6\omega)t^2+2(E-4\omega b-8\omega)}{\omega}-\frac{1+b}{2}\right)\partial_t \\
\phantom{L_2 =}{}
+\frac{2t((E-4\omega-2\omega b)^2-4\omega^2a^2)}{\omega^2}-\frac{2E(1+b)}{\omega}+4b^2+8b+5.
\end{gather*}

As was proposed in the Ansatz,  $L_1$ is diagonalized by monomials in $t$ and further the operators~$L_1$ and~$L_2$ act as automorphisms on polynomials in~$t$. Thus, we take the set $\lbrace t^k\,|\,k=0,\ldots \rbrace$ to be the basis for our representation. The three-term recurrence relation generated by the action of $L_2$ on monomials is
\begin{gather}
 \label{threeterm}
  L_2t^n=C_{n+1,n}t^{n+1}+C_{n,n}t^n +C_{n-1,n}t^{n-1},
 \end{gather}
where
\begin{gather*}
C_{n+1,n} =\frac{2(E-4\omega-2a\omega-2b\omega-4n\omega)(E-4\omega+2a\omega-2b\omega-4n\omega)}{\omega^2},\\
C_{n,n} =\frac{-16n^2\omega+(4E-16\omega(1+b))n-2E(1+b)+\omega(4b^2+8b+5)}{2\omega},\\
C_{n-1,n} =\frac{n(n+b)}{2}.
\end{gather*}

Note that for arbitrary parameters, $C_{-1,0}$ vanishes  whereas $C_{n+1,n}$ in general does not vanish for any value of~$n$ and so the  so the representation is inf\/inite dimensional, bounded below. If we make the assumption that $C_{n+1,n}$ is zero for some f\/inite integer, then we obtain quantization conditions for the parameters and a f\/inite representation of dimension $m$. In particular, we chose to solve $C_{m,m-1}=0$ for the energy value $E$ to obtain
\begin{gather}
\label{EE1} E=-2\omega (2m +  a +b),   \qquad m\in \mathbb{N}.
\end{gather}
The potential and hence the algebra relations are symmetric under the transformation $b\rightarrow -b$ and $a\rightarrow -a$ and so we could have taken the opposite sign for either $a$ or $b$ for the construction of the model and hence in the resulting quantization of the energy values.

The existence of a three-term recursion formula as in \eqref{threeterm} indicates the existence of raising and lowering operators. Indeed, they are given by the following relations
\begin{gather*}
A \equiv L_2+\frac{R}{4\omega}-\frac{L_1^2}{2\omega^2}+\frac{E}{2\omega^2}L_1-\frac12  =t\partial_t^2+(1+b)\partial_t, \\
A^\dagger \equiv L_2-\frac{R}{4\omega}-\frac{L_1^2}{2\omega^2}+\frac{E}{2\omega^2}L_1-\frac12 \\
\phantom{A^\dagger}{} =64t^3\partial_t^2-\frac{32(E-2\omega b-6\omega)}{\omega}t^2\partial_t+\frac{4((E-2\omega b-4\omega)^2-4a^2\omega)}{\omega^2} t. \end{gather*}

Here we note that if $L_1$ and $L_2$ are self-adjoint, A and $A^\dagger$ will only be mutual adjoints if $\omega$ is real, which is ref\/lective of the physical fact that the potential will be attracting in that case.
The commutation relations of the raising and lowering operators can either be determined from the quadratic algebra (\ref{E1alg1})--(\ref{E1alg2}) or directly from the model. They are
\begin{gather*}
[L_1,A] =4\omega A, \qquad [L_1,A^\dagger]=-4\omega A^\dagger,\\
[A,A^\dagger]=-\frac{4}{\omega^3}L_1^3+\frac{6E}{\omega^3}L_1^2-\frac{2}{\omega^3}\left(E^2-4\omega^2\big(a^2+b^2+2\big)\right)L_1-\frac{8E(a^2-1)}{\omega}.
\end{gather*}

We return to f\/inite dimensional model where the basis of eigenvectors for  $L_1,$ is given by $ \lbrace \phi_n(t)=k_nt^n\,|\, n=0,\ldots, m-1\rbrace$. Let us assume the existence of an inner product for which $L_1$ and $L_2$ are self-adjoint, or practically, we assume $A$ and $A^{\dagger}$ are mutual adjoints. As mentioned above,  this is equivalent to $L_1$ and $ L_2$ being self-adjoint and the constants~$a$,~$b$, and $\omega$ being real.  With this inner product, we can f\/ind the normalization for our eigenvectors using the raising and lowering operators. That is, we assume
\[
\langle At^n,t^{n-1}\rangle=\langle t^{n}, A^\dagger t^ {n-1}\rangle
\]
 and $\langle \phi_n,\phi_n \rangle=1$ to obtain the recursion relation as
 \[
 k_n^2=\frac{64(m-n)(m-n+a)}{n(b+n)}k_{n-1}^2,
 \]
 so that
\[
k_n=16^n\sqrt{\frac{(-m)_n(-m-a)_n}{n!(b)_n}}
\]
which is real so long as $a,b>0$.

From these normalization coef\/f\/icients, we can f\/ind a reproducing kernel for this Hilbert space which lies in the Hilbert space. It has the characteristic $\langle \delta(t\overline{s}),f(t)\rangle=f(s)$ and is given by $\sum \phi_n(t)\phi_n(\overline{s})=\sum k_n^2(t\overline{s})^n$ which is exactly the hypergeometric polynomial
\[
\delta (t,\overline{s})={}_2F_1\left(\begin{array}{cc} -m, &-m-a\\ b \end{array}\Bigg| \, t\overline{s}\right).
\]

Finally, it is possible to construct  an explicit function space  representation for the inner product which will make $L_1$ and $L_2$ self-adjoint. We assume the inner product is of the form $\langle f(t),g(t)\rangle=\int_{\gamma}\overline{f(t)}g(t)\rho(t,\overline{t})dtd\overline{t}$ where $\gamma$ is a path to be determined later. We can determine the weight function using the following relation,
\begin{gather*}
\langle L_1f,g \rangle=\int_{\gamma}\overline{(L_1f(t))}g(t)\rho(t,\overline{t})dtd\overline{t}=\int_{\gamma}\overline{f(t)}(L_1g(t))\rho(t,\overline{t})dtd\overline{t}=\langle f,L_1 g\rangle.
\end{gather*}
Similarly, we require $\langle Af,g\rangle=\langle f,A^{\dagger} g\rangle.$ These two integral equations  can be transformed by integration by parts into dif\/ferential equation for the weight function whose solutions are
\begin{gather*}
\rho(t\overline{t})=c_1\; {}_2F_1\!\left(\begin{array}{cc} 1+m, &m+1+a\\1- b \end{array} \Bigg|\, t\overline{t}\right) +c_2  (t\overline{t})^b \; {}_2F_1\!\left(\begin{array}{cc} 1+m+b, & m+1+a+b\\1+b \end{array}\Bigg| t\overline{t}\right).\!
\end{gather*}

Finally, we note that we can also diagonalize the operator $L_2$ in the model. The solutions of the eigenvalue equation for $L_2,$  $(L_2-\lambda)\psi_\lambda=0,$ are hypergeometric functions, ${}_2F_1$'s \cite{yellowbook}. If we assume the model is f\/inite dimension and restrict $E$ to the values in~(\ref{EE1}), then~$L_2$  becomes
\begin{gather*}
L_2 =\frac{t(8t-1)^2}{2}\partial_t^2-\frac{(8t+1)((2m+a-3)8t-b-1)}{2}\partial_t\\
\phantom{L_2 =}{} +32(m-1)(m-1+a)t-4m(b+1)-2ba-2a+2b+\frac52.
\end{gather*}  For a f\/inite dimensional irreducible representation, the hypergeometric series must terminate and we obtain a quantization relation on the eigenvalues $\lambda=-3/2-2b-2a-4k-2ba-4bk-4ak-4k^2$ and the basis functions become, for $k=1,\dots, m-1$,
\begin{gather*}
\psi_k(t) = l_k (8t+1)^{m-1-k}\; {}_2F_1\left(\begin{array}{cc} -k, &-a-k\\ 1+b \end{array}\Bigg|-8t\right)  = l_k (8t+1)^{m-1}P_k^{b,a}\left(\frac{1-8t}{1+8t}\right).
\end{gather*}
Here the $l_k$'s are normalization constants and $P_k^{b,a}$ is a Jacobi polynomial.

This model is an interesting example of the simplicity of dif\/ferential models. We can directly compute the eigenvalues for all the operators and the normalizations of the basis vectors. Also, the kernel function and integral representation of the inner product give useful tools to f\/ind expansion formulas and possible generating functions for the original quantum mechanical problem.

In the remainder of the paper, we exhibit a function space realization for a representative of each of the St\"ackel equivalence classes of second-order superintegrable systems in~2D.

\section{Models of quadratic algebras for non-degnerate second-order\\ superintegrable systems}\label{section3}

We will make use of the following conventions. On Euclidean space, the generators of the Killing vectors are,
\[
p_x=\partial_x, \qquad p_y=\partial_y, \qquad M=x\partial_y-y\partial_x.
\]
We also def\/ine the operator $p_{\pm}\equiv \partial_x\pm i\partial_y.$

The Laplacian on $E_{2,{\mathbb C}}$ in Cartesian coordinates is
\[
\Delta\equiv \partial_x^2+\partial_y^2.
\]
In complex coordinates, $z=x+iy$, $\overline{z}=x-iy,$ we have
\[
\Delta\equiv \partial_z\partial_{\overline{z}}.
\]
Or, we can take a real form in lightlike coordinates, $\nu=x$, $\zeta=iy,$ to obtain the wave operator,
\[
\Delta\equiv \partial_\nu^2-\partial_\zeta^2.
\]

For systems on the two sphere, we use the coordinates of the standard embedding of the sphere into 3 dimensional Euclidean space. We denote these $s_1$, $s_2$, $s_3$ such that $s_1^2+s_2^2+s_3^2=1.$ The basis for the Killing vectors is
\[
J_i=\sum_{i,j,k}\epsilon^{ijk} s_j\partial_{s_k}.
\]
The Laplacian on $S_{2,{\mathbb C}}$ is
\[
\Delta_{S^2}\equiv \sum_{i=1}^{3}J_i^2.
\]

The quantum algebra structures often have symmetrized terms. We def\/ine these by $\lbrace a, b \rbrace \equiv ab+ba$ and $\lbrace a,b,c\rbrace \equiv abc+acb+bac+bca+cab+cba.$  Also used is the permutation sign  function $\epsilon_{ijk}$ and $\epsilon_{ijkl}$, the completely skew-symmetric tensor on three or four variables respectively.

In the following sections, we exhibit irreducible representations of the quadratic algebras, def\/ined by relations (\ref{a1}), 
 using both dif\/ferential and dif\/ference operator realizations of the operators acting on function spaces.  Because the Hamiltonian operator commutes with all of operators, it must be a constant for any irreducible representation and so we restrict the Hamiltonian to a constant energy $H\equiv E$. We focus on models which will diagonalize operators associated with separation of variables, since these are of immediate interest for explicit solution of the physical system though additional models are given as well. Most of the models were obtained by assuming a dif\/ferential operator Ansatz except in the case of S9 which is realized as a dif\/ference operator. In this case, the operators were derived from the abstract structural relations though it is interesting to note that they also could have been obtained through the quantization of a model for the Poisson algebra for the associated classical system~\cite{KMPost2, KMPclassical}. Unless otherwise noted, these models were f\/irst exhibited in~\cite{Postthesis}.

We classify the nondegenerate systems in 2D by their St\"ackel equivalence classes and identify them by the leading order terms of the functional relation, as seen in the accompanying Table~\ref{stackelnondegen}. This classif\/ication comes from~\cite{Kress2007} and the nomenclature of the Ei's comes from~\cite{Eis}. In this section, we describe the symmetry operators, quadratic algebras and at least one model for each of the equivalence classes.
\begin{table}[t]
\caption{St\"ackel equivalence classes of non-degenerate systems in 2 dimensions.}\label{stackelnondegen}\centering

\vspace{1mm}

\begin{tabular}{|| l | l | l||}
 \hline

  \hline
  \textbf{Leading terms of Casimir relation} & \textbf{System} & \textbf{Operator models} \\ \hline
  $L_1^3\qquad \qquad \quad+f(\alpha_i,H)L_2^2$ & E2, S1& Dif\/ferential \tsep{2pt}\\
  $L_1^3 \qquad \qquad \quad+f(\alpha_i,H)L_1L_2$ & E9, E10 & Dif\/ferential\\
  $L_1^3 \qquad \qquad\quad+0$ & E15& Dif\/ferential \\
  $L_1^2L_2\qquad\quad\quad +f(\alpha_i,H)L_2^2$& E1, E16, S2, S4 & Dif\/ferential\\
  $L_1^2L_2\qquad \quad\quad +0$  & E7, E8, E17, E19& Dif\/ferential \\
  $L_1L_2(L_1+L_2)  +f(\alpha_i,H)L_1L_2$& S7, S8, S9 & Dif\/ference \\
  $0\qquad \qquad \qquad +f(\alpha_i, H)L_1L_2$ & E3, E11, E20& Dif\/ferential \\
  \hline
\end{tabular}
\end{table}

\subsection[E2: Smorodinsky-Winternitz ii]{E2: Smorodinsky--Winternitz ii}
\label{E2}

The Hamiltonian is on real Euclidean space for real constants $\omega$, $b$, $c$
\[
H=\Delta-\omega^2(4x^2+y^2)+bx+\frac{\frac14-c^2}{y}.
\]
A basis for the symmetry operators is given by $H$ and
\begin{gather*}
 L_1 = p_{x}^2-4\omega^2x^2+bx,\qquad
L_2 = \frac12\lbrace M, p_{y} \rbrace - y^2\left(\frac b4-x \omega^2\right)+\left(\frac14-c^2\right)\frac x{y^2}.
\end{gather*}
The symmetry algebra relations are, recall $R\equiv[L_1,L_2]$,
\begin{gather} \label{E2alg} [L_1,R] = -2bH+16 \omega^2 L_2+2b L_1,\\
 \left[ L_2, R \right]  =  8L_1H-6L_1^2-2H^2+2bL_2-8\omega^2 \big(1-c^2\big),\\
R^2 =  4L_1^3+4L_1H^2-8L_1^2H+16\omega^2L_2^2-4bL_2H+2b \lbrace L_1,L_2\rbrace \nonumber \\
\phantom{R^2 =}{} +16\omega^2\big(3-c^2\big)L_1-32\omega^2H-b^2\big(1-c^2\big). \label{E2algf}
\end{gather}
The algebra relations (\ref{E2alg})--(\ref{E2algf}) are realized by $H=E$ and the following operators,
\begin{gather*}
 L_1 = 4t\omega\partial_t+2\omega+\frac1{16}\frac{b^2}{\omega^2},\\
L_2 = 32\omega t^3\partial_t^2+\left(\left(16(E-6\omega)+\frac{b^2}{\omega^2}\right)t^2-\frac{b}{2\omega}t-\frac18\right)\partial_t\\
\phantom{L_2 =}{} +\frac{(16E\omega^2-64\omega^3-b^2)^2-(32\omega^3c)^2}{128 \omega^5}t +\frac{16E\omega^2-32\omega^3-b^2}{128\omega^4}.
\end{gather*}
Here, the eigenfunctions of $L_1$ are monomials and the action of $L_2$ on this basis can be represented as a three-term recurrence relation given by
\begin{gather*} 
L_2t^n=C_{n+1,n}t^{n+1}+C_{n,n}t^n +C_{n-1,n}t^{n-1},
\end{gather*}
where
\begin{gather*} C_{n+1,n} = \frac{(16E\omega^2-64\omega^3-b^2-64\omega^3n)^2-(32\omega^3 c)^2}{16\omega^2},\\
C_{n-1,n} = \frac{n}{8},\qquad
C_{n,n} = \frac{b(16E\omega^2-32\omega^3-b^2-64\omega^2n)}{128\omega^4}.
\end{gather*}

Notice, that $C_{-1,0}$ vanishes and so our representation is bounded below. Further, if there exists some~$n$ so that $C_{n+1,n}$ also vanishes then the representation becomes f\/inite dimensional. Conversely, we can assume that the representation is f\/inite dimensional to obtain quantization conditions on the energy. If we assume that representation space is $m$ dimensional and spanned by the monomials $\lbrace t^n \, |\, n=0,\ldots, m-1\rbrace,$ then the restriction $C_{m,m-1}=0$ gives the quantization of the energy values
\begin{gather*} 
 E=4\omega(m+2\epsilon c)+\frac{b^2}{16\omega^2},\qquad   m\in \mathbb{N}.
\end{gather*}
Here, we can take either $\epsilon =\pm 1$ in the energy which is consistent with the Hamiltonian depending only $c^2.$

Finally, we note that we can def\/ine raising and lowering operators in this model as,
\begin{gather*} A =L_2-\frac{R}{4 \omega}+\frac{bL_1}{4\omega^2}-\frac{bE}{4\omega^2} =\frac1{\omega^2}\partial_t,\\
 A^{\dagger} =L_2+\frac{R}{4 \omega}+\frac{bL_1}{4\omega^2}-\frac{bE}{4\omega^2}\\
 \phantom{A^{\dagger}}{}
 =t\left(64\omega t^2\partial_t^3-\frac{128\omega^3 (16E\omega^2-96\omega^3-b^2)}{64\omega^5}t\partial_t+
 \big(16E\omega^2-64\omega^3-b^2\big)^2-\big(32\omega^3c\big)^2\right).
 \end{gather*}
The raising and lowering operators obey the following commutation relations
\begin{gather*} [L_1,A] =-4\omega A,  \qquad  [L_1,A^\dagger]=4\omega A^\dagger,\\
[A,A^\dagger]=-\frac3{\omega} L_1^2+\frac{32\omega^2+b}{8\omega^3}L_1-\frac{1}{8\omega^3}\big(8E^2 \omega^2+b^2E+32\omega^4\big(1-c^2\big)\big).
\end{gather*}

\subsection{E10}
\label{E10}

The Hamiltonian for this system is given by, with $z=x+iy$, $\overline{z}=x-iy$
\[
H=\Delta +\alpha\zbar +\beta\left(z-\frac32\zbar^2\right)+\gamma\left(z\zbar-\frac12\zbar^3\right).
\]
A basis for its symmetry operators is given by $H$ and
\begin{gather*} L_1 =p_{-}^2+\gamma \zbar^2+2\beta\zbar,\\
L_2 =2i\lbrace M,p_{-} \rbrace +p_+^2-4\beta z\zbar-\gamma z \zbar^2-2 \beta \zbar^3-\frac{3}{4} \gamma \zbar^4+\gamma z^2+\alpha\zbar^2+2\alpha z.
\end{gather*}
The algebra relations are given by
\begin{gather*} 
[R,L_1] =-32\gamma L_1-32\beta^2, \qquad
 [R,L_2] =96L_1^2-128\alpha L_1+32\gamma L_2+64\beta H+32\alpha^2,\\
 R^2 =64L_1^3+32\gamma\lbrace L_1, L_2 \rbrace -128\alpha L_1^2-64\gamma H^2-128\beta H L_1+64\beta^2L_2+64\alpha^2L_1\\
\phantom{R^2 =}{} -128\beta\alpha H-256\gamma^2.
 \end{gather*}
This algebra can be transformed into  the Lie algebra $sl_2$ by using the following invertible transformation
\begin{gather*} K_1 =L_1+\frac{\beta^2}{\gamma},\qquad
 K_2 =L_2+\frac1{\gamma}L_1^2- \frac{\beta^2+2\alpha\gamma}{\gamma^2}L_1+\frac{2\beta}{\gamma}H+\frac{(\alpha \gamma+\beta^2)^2}{\gamma^3}.
 \end{gather*}
In this basis, the algebra relations reduce to a Lie algebra
\begin{gather}\label{E10alg}   [R,K_1]=-32\gamma K_1,\qquad   [R,K_2]=32\gamma K_2, \qquad [K_1,K_2]=R,\\
\label{E10algf}   R^2=32\gamma \lbrace K_1,K_2\rbrace-64\gamma H^2-\frac{128\beta(\alpha\gamma+\beta^2)}{\gamma}H-\frac{64(\beta^6+4\gamma^4+\alpha^2\beta^2\gamma^2+2\alpha \beta^4\gamma)}{\gamma^3}.\!\!\!
\end{gather}

The algebra relations (\ref{E10alg}), (\ref{E10algf}) are realized by $H=E$ and   the following operators,
\begin{gather*}
K_1 =16\gamma\partial_t,\qquad
K_2 =t^2\partial_t+\left(1+\frac{\sqrt{-\gamma}(\gamma^2E+\alpha\beta\gamma+\beta^3)}{2\gamma^3}\right)t,\\
R \equiv[L_1,L_2]=16\gamma\left(2t\partial_t+\left(1+\frac{\sqrt{-\gamma}(\gamma^2E+\alpha\beta\gamma+\beta^3)}{2\gamma^3}\right)\right).
\end{gather*}

The operator $R$ is diagonalized by monomials $t^n$ and in this basis $K_1$ and $K_2$ act as lowering and raising operators respectively. Note that $K_1$ annihilates constants and so the representation is bounded below. On the other hand, for a f\/inite dimensional representation, we require that there exist some integer $m$ such that
\[
 K_2t^{m-1}=\left(m-1+\left(1+\frac{\sqrt{-\gamma}(\gamma^2E+\alpha\beta\gamma+\beta^3)}{2\gamma^3}\right)\right)t^{m}=0.
  \]
  This leads to the quantization condition on the energy
\begin{gather}\label{EE10}
E=2i\sqrt{\gamma}m-\frac{\alpha\beta}{\gamma}-\frac{\beta^3}{\gamma},\qquad m\in\mathbb{N}.
\end{gather}
 Notice that in order to obtain a real energy value we require that all the parameters be real and that $\gamma<0.$

 It is also possible to diagonalize a linear combination of $K_1$ and $K_2.$ For example,
 the solutions of the eigenfunction equation
 \[
  (K_1+K_2-\lambda)\Psi=0
  \]
  are
\[
\Psi=\big(16\gamma+t^2\big)^{\frac{\sqrt{-\gamma}(\gamma^2E+\alpha\beta\gamma+\beta^3)}{4\gamma^3}}
 \exp\left(\frac{\lambda}{4\sqrt{\gamma}}\arctan\left(\frac{t}{4\sqrt{\gamma}}\right)\right)
 \]
 which, for f\/inite dimensional representations where $E$ is restricted to the value in (\ref{EE10}), give quantization conditions on the eigenvalues $\lambda$. That is, if we require that $\Psi$ be a polynomial in $t$ of degree less that $m$ we obtain a complete set of eigenfunctions given by
 \begin{gather*} (K_1+K_2)\Psi_n =\lambda_n\Psi_n, \qquad n=0,\ldots, m-1,\\
   \Psi_n =\big(4\sqrt{-\gamma}+t\big)^n\big(4\sqrt{-\gamma}-t\big)^{m-n-1}, \qquad
   \lambda_n  =4\sqrt{-\gamma}(m-2n-1).
   \end{gather*}

 Finally, we can return to the original basis of $L_1$ and $L_2$. In the model,
 \begin{gather*}
 L_1=16\gamma\partial_t-\frac{\beta^2}{\gamma},\\
L_2 =256\gamma\partial_t^2+\left(t^2+32\alpha+\frac{48\beta^2}{\gamma}\right)\partial_t\\
\phantom{L_2 =}{}
+\left(1+\frac{\sqrt{\gamma}(\gamma^2E+\alpha\beta\gamma+\beta^3)}{2\gamma^3}\right)t
-\frac{2\beta\gamma^2E+\alpha^2\gamma^2+4\alpha\beta^2\gamma+3\beta^4}{\gamma^3}.
\end{gather*}

\subsection{E15}\label{E15}

Here
\[
H=\Delta +h(\zbar),
\]
where the potential is an arbitrary function of $\zbar.$  A basis for the  symmetry operators is
\[
L_1=p_{-}^2, \qquad L_2=2i\lbrace M,p_{-} \rbrace +i\int \zbar \frac{dh}{d\zbar}  d\zbar.
\]
The only nonzero algebra relation is $ [L_1,L_2]=iL_1$. This system is unique among all 2 dimensional  superintegrable systems in that the symmetry operators are not functionally  linearly independent and do not correspond to multiseparability.  The only separable system is determined by diagonalizing~$L_1$, essentially $z$, $\zbar$,  and this coordinate system is not orthogonal.
A~model is
\[
L_1= \frac{d}{dt}+ a,\qquad L_2= it\frac{d}{dt}+iat,
 \]
 but the irreducible representations of the algebra yield no spectral information about~$H$. Again, since this algebra is a Lie algebra, the above model is derivative of one found in \cite{Miller68}.

\subsection[E1, Smorodinsky-Winternitz i]{E1, Smorodinsky--Winternitz i}
\label{E1}
This is the system considered in Section~\ref{section2} and for completeness we recall the results here.  The Hamiltonian for the system is
\[
 H=\Delta -\omega^2\big(x^2+y^2\big)+\frac{\frac14-a^2}{x^2}+\frac{\frac14-b^2}{y^2}.
 \]
The remaining generators of the symmetry algebra are
\[
L_1=\partial_x^2+\frac{1/4-a^2}{x^2}-\omega^2x^2, \qquad L_2=M^2+\frac{(1/4-a^2)y^2}{x^2}+\frac{(1/4-b^2)x^2}{y^2}.
\]
The algebra relations are
\begin{gather}\label{E1alg}  \left[R,L_1\right] =8L_1^2-8HL_1+16\omega^2L_2-8\omega^2,\\
\left[R,L_2\right] =8HL_2-8\lbrace L_1, L_2 \rbrace+\big(12-16a^2\big)H+\big(16a^2 +16b^2-24\big)L_1, \\
 R^2 =8H\lbrace L_1,L_2\rbrace-\frac{8}{3}\lbrace L_1,L_1, L_2\rbrace +16\omega^2L_2^2+16\big(a^2-1\big)H^2+\left(16a^2+16b^2-\frac{200}3\right)L_1^2\nonumber\\
\phantom{R^2 =}{}
-\left(32a^2-\frac{200}3\right)HL_1-\frac{176\omega^2}{3}L_2-\frac{4\omega^2}{3}\big(48a^2b^2-48a^2-48b^2+29\big)
.\label{E1algf}
\end{gather}
The algebra relations (\ref{E1alg})--(\ref{E1algf}) are realized by $H=E$,  and   the following operators
\begin{gather*}
L_1 =-4\omega t \partial_t+E-2\omega(1+b),\\
L_2 =\frac12t(8t+1)^2\partial_t^2-\left(\frac{16(E-2\omega b-6\omega)t^2+2(E-4\omega b-8\omega)}{\omega}-\frac{1+b}{3}\right)\partial_t\\
 \phantom{L_2 =}{}
 +\frac{2t((E-4\omega-2\omega b)^2-4\omega^2a^2)}{\omega^2}-\frac{2E(1+b)}{\omega}+4b^2+8b+5.
 \end{gather*}

As described in the previous section, the eigenvalues for $L_1$ are monomial and the eigenfunctions for $L_2$ are hypergeometric functions which reduce to Jacobi polynomials for f\/inite representations. The representation become f\/inite dimensional under the quantization of energy $E=-2\omega (2m +  a +b),$ \eqref{EE1}.

\subsection{E8}
\label{E8}

The Hamiltonian is
\[
H=\Delta +\frac{\alpha z}{\zbar^3}+\frac{\beta}{\zbar^2}+\gamma z\zbar.
\]
A basis for the symmetry operators  is given by $H$ and
\[
L_1=p_{-}^2+\frac{\zbar^4-\alpha}{\zbar^2}, \qquad L_2=M^2+\beta\frac{z}\zbar +\alpha\frac{z^2-\zbar^2}{\zbar^2}.
\]
The algebra relations are
\begin{gather} \label{E8alg} [R,L_1] =8{L_1}^2+32 \alpha \gamma,\qquad
 [R,L_2] =-8\{L_1,L_2\}+8bH-16( \alpha+1)L_1,\\
 R^2 =-\frac{8}{3}\lbrace L_1,L_1,L_2\rbrace -\left(16 \alpha+\frac{176}{3}\right)L_1^2 +16 \alpha H^2
-64 \alpha \gamma L_2+16 \beta L_1H\nonumber\\
\phantom{R^2 =}{} -64 \gamma \alpha ^2-16 \gamma \beta ^2+\frac{64}{3} \alpha \gamma.\label{E8algf}
\end{gather}
The algebra relations (\ref{E8alg}), (\ref{E8algf}) are realized by $H=E$ and   the following operators
\begin{gather*} L_1 =2\sqrt{ -\alpha \gamma}t,\qquad
L_2 =-4\big(t^2-1\big)\partial_t^2+\left(\left(\frac{2\beta}{\sqrt{\alpha}}-8\right)t +\frac{2E}{\sqrt{-\gamma}}\right)\partial_t-\left(1+\frac{\beta}{2\sqrt{\alpha}}\right)^2-\alpha.
\end{gather*}
Here,  the eigenfunctions of $L_2$ are hypergeometric functions which restrict to Jacobi polynomials under quantization of eigenvalues
\[ 
L_2\Psi_n=\lambda_n\Psi_n, \qquad \lambda_n=-4n^2+\frac{(8\beta\sqrt{\alpha}-16\alpha)}{4\alpha}n-\frac{4\alpha^2+4\alpha+4\beta\sqrt{\alpha}-\beta^2}{4\alpha}.
\]
The eigenfunctions are given by
\[\Psi_n=l_nP_n^{a,b}(-t),\qquad a=\frac{E}{4\sqrt{-\gamma}}-\frac{\beta}{4\sqrt{\alpha}}, \qquad b=-\frac{E}{4\sqrt{-\gamma}}-\frac{\beta}{4\sqrt{\alpha}},
\]
where $l_n$ is a normalization constant. The action of $L_1$ in this model is via multiplication by the variable $t$ and gives a three-term recurrence formula.  Note that for some quantization of  the energy  the eigenfunctions become singular for all $n\ge m$. This occurs under the quantization condition $m\in \mathbb{N}$ with energy eigenvalues
\[
 E=2\sqrt{-\gamma}\left(2m+2\pm\frac{\beta}{2\sqrt{\alpha}}\right), \qquad   m\in\mathbb{N}.
 \]

\subsection{S9: the generic system in 2D}
\label{S9}

The Hamiltonian is
\[
 H=\Delta_{S^2}+\frac{\frac14-a^2}{s_1^2}+\frac{\frac14-b^2}{s_2^2}+\frac{\frac14-c^2}{s_3^2}.
\]

A basis for the symmetry operators is
\begin{gather*} L_1=J_3^2+\left(\frac14-a^2\right)\frac{s_1^2}{s_2^2}+\left(\frac14-c^2\right)\frac{s_2^2}{s_1^2},\qquad
 L_2=J_1^2+\left(\frac14-a^2\right)\frac{s_3^2}{s_2^2}+\left(\frac14-b^2\right)\frac{s_2^2}{s_3^2},\\
 H=L_1+L_2+L_3+\frac34-a^2-b^2-c^2.
 \end{gather*}

The structure equations can be put in the symmetric form using the following identif\/ications
\begin{gather}
a_1=\frac14-c^2,\qquad  a_2=\frac14-a^2,\qquad a_3=\frac14-b^2,\nonumber\\
  \label{S9alg}
  [L_i,R] =\epsilon_{ijk}\left(4\{L_i,L_k\}-4\{L_i,L_j\}- (8+16a_j)L_j + (8+16a_k)L_k+ 8(a_j-a_k)\right),\\
 R^2 =\frac83\{L_1,L_2,L_3\} -(16a_1+12)L_1^2 -(16a_2+12)L_2^2  -(16a_3+12)L_3^2\nonumber\\
\phantom{R^2 =}{} +\frac{52}{3}(\{L_1,L_2\}+\{L_2,L_3\}+\{L_3,L_1\})+ \frac13(16+176a_1)L_1+\frac13(16+176a_2)L_2 \nonumber\\
\label{S9algf}
\phantom{R^2 =}{}
+ \frac13(16+176a_3)L_3 +\frac{32}{3}(a_1+a_2+a_3)+48(a_1a_2+a_2a_3+a_3a_1)+64a_1a_2a_3.
\end{gather}

We can obtain $L_1$ in the model by using the following dif\/ference operator, based upon the Wilson polynomial algebra. This model is unique in that the energy values $E$ and eigenvalues of the operators were used in determining the model. We def\/ine the coef\/f\/icients $\alpha$, $\beta$, $\gamma$ and $\delta$ as
\begin{gather*}
 \alpha=  -\frac{a+c+1}{2}-\mu  ,\!\qquad \beta=\frac{a+c+1}{2},\!\qquad  \gamma=\frac{a-c+1}{2}  ,\!\qquad \delta=\frac{a+c-1}{2}+b+\mu+2,
 \end{gather*}
and use dif\/ference operators
\begin{gather*}
T^AF(t)=F(t+A),\qquad
\tau=\frac{1}{2t}(T^{1/2}-T^{-1/2}),\\
  \tau^*=\frac{1}{2t}\big[(\alpha+t)(\beta+t)(\gamma+t)(\delta+t)T^{1/2}-(\alpha-t)(\beta-t)(\gamma-t)(\delta-t)T^{-1/2}\big].
\end{gather*}
The algebra relations (\ref{S9alg}), (\ref{S9algf}) are realized by the following operators
\begin{gather*}
L_3 =-4t^2+a^2+c^2,\qquad
L_1 =-4\tau^*\tau-2(a+1)(b+1)+\frac12,\\
H =E, \qquad E\equiv-\frac14 (4\mu+2a+2b+2c+5)(4\mu+2a+2b+2c+3)+\frac32 -a^2-b^2-c^2.
\end{gather*}

The model realizes the algebra relations for arbitrary complex $\mu$ and restricts to a f\/inite dimensional irreducible representation when $\mu=m\in \mathbb{N}.$ In this model, we obtain spectral resolution of $L_3$ with delta functions as eigenfunctions. The eigenfunctions of $L_1$ are Racah polynomials in the f\/inite dimensional case and Wilson polynomials for the inf\/inite dimensional, bounded below case. This model was f\/irst published in~\cite{KMPost} where the model was worked out in full generality including the normalizations and the weight functions. It has also been recently been extended to the 3D analog in~\cite{KMPost2010}.

\subsection[E20: Smorodinksy-Winternitz iii]{E20: Smorodinksy--Winternitz iii}
\label{E20}

The Hamiltonian is
\[
H=\Delta +\frac{4}{\sqrt{x^2+y^2}}\left(\alpha+\beta\frac{\sqrt{\sqrt{x^2+y^2}+x}}{\sqrt{x^2+y^2}}
+\gamma\frac{ \sqrt{\sqrt{x^2+y^2}-x}}{\sqrt{x^2+y^2}}\right).
\]
In parabolic coordinates $(x',y')$, with $x'=\sqrt{x^2+y^2}+x$, $y'=\sqrt{x^2+y^2}-x$, the Hamiltonian can be written as
\[
 H=\frac{1}{x'^2+y'^2}\big(\partial_{x'}^2+\partial_{y'}^2\big) +\frac{4(\alpha-\beta x'-\gamma y')}{x'^2+y'^2},
 \]
The integrals are
\begin{gather*}  L_1 =\frac{1}{x'^2+y'^2}\left(y'^2\partial_{x'}^2-x'^2\partial_{y'}^2-2\alpha\big(x'^2-y'^2\big)-4\beta x'y'^2+4\gamma x'^2y'\right),\\
 L_2 =\frac{1}{x'^2+y'^2}\left(-x'y'\big(\partial_{x'}^2+\partial_{y'}^2\big)+\big(x'^2+y'^2\big)\partial_{x'}\partial_{y'}-4\alpha x'y'+2\big(x'^2-y'^2\big)(y'\beta-x'\gamma)\right).
 \end{gather*}
The algebra relations are given by,
 \begin{gather}\label{E20alg}   [R,L_1] =-4L_2H+16 \beta \gamma,\qquad
 [R,L_2] =4L_1H-8\big(\beta^2-\gamma^2\big),\\
\label{E20algf}  R^2 =4L_1^2H+4L_2^2H+4H^2-16\alpha^2H+16\big(\gamma^2-\beta^2\big)L_1-32\beta\gamma L_2-32\alpha^2\big(\beta^2+\gamma^2\big).
\end{gather}
Notice, these restrict to a Lie algebra on a constant energy surface, $H=E$, and so the model described below is not new but, for example, can be derived from those given in~\cite{Miller68}.

The algebra relations (\ref{E20alg}), (\ref{E20algf}) are realized by $H=E$ and   the following operators,
\begin{gather*} L_1 =-2\sqrt{E}t\dop -\sqrt{E}+2\alpha+4\frac{\beta^2}{E},\\
L_2 =-\frac{1}{2}Et^2\dop+2\dop-\frac{1}{2}tE+\sqrt{E}at+\left(\frac{\beta^2}{\sqrt{E}}+\frac{\gamma^2}{\sqrt{E}}\right)t-4\frac{\beta\gamma}{\sqrt{E}}. \end{gather*}

Here $L_1$ is diagonalized by monomials $t^n$ and the action of $L_2$ on monomials is given by the following three-term recursion formula
\[ 
L_2t^n=C_{n+1,n}t^{n+1}+C_{n,n}t^n +C_{n-1,n}t^{n-1}
\]
with
\begin{gather*} C_{n+1,n} =-(n+1)E+\frac{\alpha}{\sqrt{E}}+\frac{\beta^2+\gamma^2}{\sqrt{E}},\qquad
C_{n,n} =\frac{4\beta\gamma}{E},\qquad  C_{n-1,n} =2n.
\end{gather*}

Notice, that $C_{-1,0}$ vanishes for arbitrary parameters and so our representation is bounded below. On the other hand, if we require that the representation be f\/inite dimensional, say of dimension $m$, we obtain the restriction
\begin{gather} \label{EE8}
-mE+2\sqrt{E}a+\frac{\beta^2}{\sqrt{E}}+\frac{\gamma^2}{\sqrt{E}}=0, \qquad m\in \mathbb{N} ,
\end{gather}
which gives the quantized restrictions on the energy.

The eigenfunctions for $L_2$ are given by
\begin{gather*}
 \Psi=\big(Et^2-4\big)^{E^{-\frac32}(E\alpha+\beta^2+\gamma)-\frac12} \exp\left(\frac{\lambda E-4\beta \gamma}{E^{\frac32}}{\rm arctanh}
 \left(\frac{\sqrt{E}t}{2}\right)\right),
 \end{gather*}
where  $L_2\Psi =\lambda\Psi.$ If we assume that the model is f\/inite dimensional, i.e.\ that the energy value~$E$ satisf\/ies~\eqref{EE8}, then the eigenvalues are quantized as
\begin{gather*}
 \lambda= \lambda_n\equiv \frac{4\beta\gamma}{E}+(m-1-2n)\sqrt{E}
 \end{gather*}
and there exists a complete set of eigenfunctions $\Psi_n$ satisfying
\begin{gather*}
 L_2\Psi_n=\lambda_n\Psi_n, \qquad \Psi_n=\big(\sqrt{E}t-2\big)^{m-n-1}\big(\sqrt{E}t-2\big)^{n}.
 \end{gather*}

This model admits raising and lowering operators,
\begin{gather*}
A =L_2+\frac{R}{2\sqrt{E}}-\frac{\beta\gamma}{E} =4\partial_t,\\
A^\dagger =L_2-\frac{R}{2\sqrt{E}}-\frac{\beta\gamma}{E}
 =-Et^2\partial_t+\frac{2\alpha E-E^{\frac32}+2\beta^2+2\gamma^2}{\sqrt{E}}t,
 \end{gather*}
which satisfy the commutation relations
\begin{gather*} [A,L_1] =-2\sqrt{E}A,  \qquad [A^\dagger, L_1]=2\sqrt{E}A^\dagger,\\
[A,A^\dagger] =-2L_1^2+\frac{\beta^2-\gamma^2}{E}L_1-2E+8\alpha^2+\frac{16\alpha(\beta^2+\gamma^2)}{E}+\frac{32\beta^2\gamma^2}{E^2}.
\end{gather*}

\section{2D degenerate systems}\label{section4}

Next, we consider degenerate systems whose symmetry algebra includes a f\/irst-order integral,~$X$. As described above, this requires that the potential depend on only one parameter, not including the trivial additive constant. In these systems, the symmetry algebra is def\/ined by the operators $H$, $L_1$, $L_2$ and~$X$. It is always possible to rewrite the commutator $R=[L_1,L_2]$ as a polynomial in the other operators of maximal degree 3 in $X$ and 1 in $L_1$, $XL_1$, $L_2$, $XL_2$, $H $  and~$XH.$  In these systems, the commutation relations are in terms of~$X$ instead of~$R.$ That is, the def\/ining relations are
\begin{gather*}
[L_1,X]=P_1\big(L_1,L_1,X^2,X,H\big),\qquad [L_1,X]=P_2\big(L_1,L_1,X^2,X,H\big),\\
 [L_1,L_2]=Q\big(X^3,XL_1,XL_2,XH,L_1,L_2,H,X\big).
 \end{gather*}
Here, the $P_i$'s and $Q$ are linear in the arguments. Furthermore, the functional relation is no longer in terms of $R^2$ but instead a fourth-order identity.

We begin with the table of the equivalence classes of degenerate systems; there are exactly 6 degenerate systems in 2 dimensions.
\begin{table}[!h]\centering
\caption{St\"ackel equivalence classes of degenerate systems which admit a Killing vector $X$, in 2D.}\label{stackeldegen}
\vspace{1mm}

\begin{tabular}{|| l | l|l||}
\hline
\textbf{Leading order terms}& \textbf{System}& \textbf{Operator models}\\
\hline
$0\qquad+L_1L_2 \qquad \quad +AX^2$& E3, E18 &Dif\/ferential  \tsep{2pt}\\

$X^4\quad +L_1L_2$ & S3, S6& Dif\/ferential and Dif\/ference \\

$X^4\quad+X^2L_1+L_2^2\quad+0$ & E12, E14& Dif\/ferential\\

$0\qquad+X^2L_1+L_2^2\quad+AL_1$ & E6, S5& Dif\/ferential\\

$X^4\quad+L_1^2$ & E5& Dif\/ferential\\

$0\qquad+X^2L_1 \qquad \quad +L_2$ & E4, E13 &Dif\/ferential\\

\hline
\end{tabular}
\end{table}

\subsection{E18: the Coulomb system in 2D}
\label{E18}

This system is def\/ined by the Hamiltonian
\[
H=\Delta+\frac \alpha {\sqrt{x^2+y^2}}.
\]
A basis for the symmetry operators is formed by $H $, $X=M$ and
\[
L_1=\frac12\lbrace M,p_x\rbrace -\frac{\alpha y}{2\sqrt{x^2+y^2}}, \qquad L_2=\frac12\lbrace M,p_y\rbrace -\frac{\alpha x}{2\sqrt{x^2+y^2}}.
\]

The symmetry algebra is def\/ined by the following relations
\begin{gather} \label{E18alg} [L_1,X] =L_2, \qquad [L_2,X]=-L_1,\qquad  [L_1,L_2] =HX,\\
\label{E18algf}  L_1^2 +L_2^2-HX^2+\frac{H-\alpha^2}4=0.
\end{gather}

We can change basis so that the algebra is in the standard form of the Lie algebra $sl_2$. With the substitutions
$A=L_1+iL_2$, $ A^\dagger=L_1-iL_2$ the algebra relations (\ref{E18alg}), (\ref{E18algf}) become,
\begin{gather} \label{E18alg'}
[A,X] =-iA, \qquad [A^\dagger,X]=iA^\dagger, \qquad
 [A,A^\dagger] =2iHX,\\
\label{E18algf'} \lbrace A,A^\dagger\rbrace -2HX^2+\frac{H}2-\frac{\alpha^2}2=0.
\end{gather}

The Lie algebra def\/ined by (\ref{E18alg'}), (\ref{E18algf'}) is realized by $H=E$ and the following operators
\begin{gather*}
X =-i\left(t\partial_t+\frac{1}{2}+\frac{\alpha}{2\sqrt{E}}\right),\qquad
A =\frac{E}{4}\partial_t,\qquad
A^\dagger =-4t\left(t\partial_t+1+\frac{\alpha}{\sqrt{E}}\right).
\end{gather*}

The operator $X$ is diagonalized by monomials $t^n$ and for arbitrary $E$ and $\alpha$ the representation is inf\/inite and bounded below. Finite dimensional representation occur if $A^\dagger$ annihilates a~monomial. That is
\[
A^\dagger t^{m-1}=-\frac{4(E m-\alpha \sqrt{E})}{E} t^m=0
\]
when the energy value $E$ takes the values
\[ 
E=\frac{\alpha^2}{m^2}, \qquad m\in \mathbb{N}.
\]
 Again, since this algebra is a Lie algebra the model employed is not new.

\subsection{S3}
\label{S3}

This system is defined by the Hamiltonian
\[
H=\Delta_{S^2}+\frac{\frac14-a^2}{s_3^2}.
\]
A basis for the symmetry operators is formed by $H$, $X=J_3$ and
\[
 L_1=J_1^2+\frac{(\frac14-a^2) s_2^2}{s_3^2}, \qquad L_2=\frac12(J_1J_2+J_2J_1)-\frac{(\frac14-a^2) s_1s_2}{s_3^2} .
\]
The algebra relations are given by
\begin{gather} \label{S3alg}  [L_1,X] =2L_2,
\qquad [L_2,X] =-X^2-2L_1+H+a^2-\frac14,
\\
 [L_1,L_2]=-\lbrace L_1,X \rbrace+\big(2a^2-1\big)X,\\
\frac1{6}\lbrace L_1,X,X\rbrace -HL_1+L_2^2+L_1^2-\left(a^2-\frac{7}{6}\right)X^2-
\left(a^2+\frac5{12}\right)L_1\nonumber\\
\qquad {}+\frac H6-\frac{5}{24}\big(4a^2-1\big)=0. \label{S3algf}
\end{gather}
The algebra relations (\ref{S3alg})--(\ref{S3algf}) are realized by $H=E$ and the following operators
\begin{gather*}
X =2it\partial_t+ic_0,\\
L_1 =t(t+1)^2 \partial_t ^2+(t+1)(c_1t+c_1+2 c_0-1) \partial_t \\
\phantom{L_1 =}{} +\left(c_1^2+\frac{3 c_0^2}{2}-2c_0 c_1+3c_0-3c_1
-\frac{a^2}{2}+\frac{E}{2}+\frac{19}{18}\right)t +\frac{a^2+c_0^2+E}{2}-\frac18, \\
L_2 =-i(t^3-t)\partial_t^2-i\left(c_1t^2+2+2c_0-c_1\right)\partial_t\\
\phantom{L_2 =}{} -i\left(c_1^2+\frac{3c_0^2}2-2c_0c_1+3c_0-3c_1-\frac{a^2}{2}+\frac{E}2+\frac{19}{18}\right) t,
 \end{gather*}
where, for compactness of the equations, we have chosen constants $c_0$, $c_1$ as
\begin{gather*}c_0 =a-1+\frac{\sqrt{4E-1}}{2},\\
c_1 =\frac34\big(2a+\sqrt{4E-1}\big)+\frac14\sqrt{4a^2+16a-12E-13-4(a-2)\sqrt{4E-1}}.
\end{gather*}
In the model, $X$ is diagonalized with monomials and both $L_1$ and $L_2$ have a three-term recurrence relation in this basis given by
\begin{gather}
\label{l1s3}
L_1t^n=C_{n+1,n}t^{n+1}+C_{n,n}t^n+C_{n-1,n}t^{n-1},\\
\label{l2s3}
L_2t^n=iC_{n+1,n}t^{n+1}-iC_{n-1,n}t^{n-1},
\end{gather}
with
\begin{gather*}
C_{n+1,n} =n^2+(c_1-1)n+c_1^2+\frac{3c_0^2}{2}-2c_0c_1-3c_1+3c_0-\frac{a^2+E}{2}+\frac{19}{8},\\
C_{n,n} =2n^2+2nc_0+\frac{c_0^2+E+a^2}{2}-\frac18,\qquad
C_{n-1,n} =n(n-c_1+2c_0+1).
\end{gather*}
Note that, for arbitrary parameters, $C_{-1,0}$ vanishes and so the model is bounded below. On the other hand, if we require that the representation space be f\/inite dimensional, i.e.\ $C_{m,m-1}=0$, we obtain the quantization condition
\begin{gather}
 \label{ES3} E=-(m-a)^2+\frac14,
  \qquad m\in \mathbb{N}.
  \end{gather}

As can be seen directly form the three-term recurrence formulas \eqref{l1s3} and \eqref{l2s3}, there exist raising and lowering operators given by
\begin{gather*}
 A^\dagger= L_1+iL_2+\frac12\left(X^2-E+\frac14-a^2\right),\qquad
A= L_1-iL_2+\frac12\left(X^2-E+\frac14-a^2\right).
\end{gather*}
The symmetry algebra of the raising and lowering operators is given by
\begin{gather*}
 [A,X] =2iA,\qquad   [A^\dagger,X]=-2iA^\dagger,\\
  \lbrace A,A^\dagger \rbrace -\frac 1 2 X^4+X^2\left(E-a^2+\frac{11}{4}\right)+\frac 1 {32}\big(4E+4a^2+8a+3\big)\big(E+4a^2-8a+3\big)=0.
  \end{gather*}

There is also a dif\/ference operator model associated with the spectral resolution of~$L_1$. The operators are formed from the operators~$T^k$ def\/ined as $T^kf(t)=f(t+k)$. When restricted to f\/inite dimensional representations, i.e.\ the energy $E$ takes the values as in~\eqref{ES3}, the algebra relations (\ref{S3alg})--(\ref{S3algf}) are realized by
\begin{gather*}
L_1 =-t^2+a^2-\frac14,\\
-iX =\frac{(1/2-a-t)(m+a-1/2-t)}{2t}T^1-\frac{(1/2-a+t)(m+a-1/2+t))}{2t}T^{-1},\\
L_2 =\frac{(1-2t)(1/2-a-t)(m+a-1/2-t)}{4t}T^1\\
\phantom{L_2 =}{} +\frac{(1+2t)(1/2-a+t)(m+a-1/2+t))}{4t}T^{-1}.
\end{gather*}
The eigenfunctions of $L_1$ are delta functions and the eigenfunctions of~$L_2$ are dual Hahn polynomials.  This model was f\/irst published in~\cite{KMPost2} where the model was worked out in full generality including the normalizations and the weight functions.

\subsection{E14}
\label{E14}

The Hamiltonian is, with $z=x+iy$, $\overline{z}=x-iy,$
\[
H=\Delta +\frac{\alpha}{\overline{z}^2}.
\]
A basis for the symmetry operators is formed by $H, $ $X=p_-$ and
\[
L_1=\frac12\lbrace M,p_{-} \rbrace+\frac{\alpha}{i\overline{z}}, \qquad L_2=M^2+\frac{\alpha z}{\overline{z}},
\]
The symmetry algebra is
\begin{gather}\label{E14alg}   [X,L_1] =iX^2,\qquad   [X, L_2]=2iL_2,\qquad
[L_1,L_2] =i\lbrace X,L_2 \rbrace +\frac{i}{2}X,\\
\label{E14algf} L_1^2 -\frac12\lbrace L_2,X^2\rbrace +\alpha H-\frac{5}{4}X^2=0.
\end{gather}
The algebra relations (\ref{E14alg}), (\ref{E14algf}) are realized by $H=E$ and the following operators
\begin{gather*}
 X=\frac1{t}, \qquad L_1=i\partial_t, \qquad L_2=-t^2\partial_t^2-2t\partial_t+\alpha Et^2-\frac1{4}.
 \end{gather*}
Eigenfunctions for $L_2$ are Bessel functions while eigenfunctions of $L_1$ are exponentials both with continuous spectrum. This model has no obvious f\/inite dimensional restrictions.

\subsection{E6}
\label{E6}

The Hamiltonian is
\[
H=\Delta + \frac{\frac14-a^2}{x^2}.
\]
A basis for the symmetry operators is formed by $H$, $X=p_y$ and
\[
L_1=\frac12 \lbrace M,p_{x} \rbrace -\frac{(\frac14-a^2)y}{x^2}, \qquad L_2=M^2+\frac{(\frac14-a^2)y^2}{x^2},
\]
The algebra relations are given by
\begin{gather} \label{E6alg}[L_1,X] =H-X^2,\qquad    [L_2,X]=2L_1,\qquad
[L_1,L_2] =\lbrace X,L_2 \rbrace +(1-2a)X,\\
\label{E6algf} L_1^2   -H L_2-2 \{ L_1,X\} +\frac{H}2+\left(\frac12-a^2\right)X^2=0.
\end{gather}
The algebra relations (\ref{E6alg}), (\ref{E6algf}) are realized by $H=E$ and the following operators
\begin{gather*}
X =\sqrt{E}\big(t^2\partial_t+(a+1)t+1\big),\\
L_1 =-\sqrt{E}\big(t^3\partial_t^2+((2a+3)t+2)t\partial_t+(a+1)^2t+a+1\big),\\
L_2 =-t^2\partial_t^2-2((a+1)t+1)\partial_t-a-\frac12.
\end{gather*}
Note that the operators $X$, $L_1$ and $L_2$ act on monomials $t^n$ via   three-term recurrence relations
\begin{gather*} Xt^n =\sqrt{E}(n+a+1)t^{n+1}+\sqrt{E}t^n,\\
L_1t^n =-\sqrt{E}(n+a+1)^2t^{n+1}-\sqrt{E}(2n+a+1)t^{n},\\
L_2t^n =-\left(n^2+(n+2a)+a+\frac12\right)t^n-2nt^{n-1}.
\end{gather*}

Thus, the model can be realized as a bounded below representation using the set of monomial $\lbrace t^n\,|\, n\in \mathbb{N} \rbrace.$ The representation will become f\/inite dimensional, with dimension~$m$ if the parameter~$a$ is restricted to quantized values $a=-m.$ Notice, a constant is quantized instead of the energy which does not make sense in the physical system, since the constants should be given by the system. However,  under a St\"ackel transform the constant and the energy will be interchanged so our model is giving quantization levels of the energy of a St\"ackel equivalent system; in this case, S5. The eigenfunctions of~$L_1$ are gauge equivalent to Laguerre polynomials with eigenvalues $\sqrt{E}(2n-m+1).$ The eigenfunctions of $L_2$ are gauge equivalent to generalized Laguerre polynomials with eigenvalues $m^2-k^2+k-\frac12$.

\subsection{E5}
\label{E5}

The Hamiltonian is
\[
H=\Delta +\alpha x.
\]
A basis for the symmetry operators is formed by $H, $ $X=p_y$ and
\[
L_1=p_xp_y+\frac1{2} \alpha y, \qquad L_2=\frac1{2}\lbrace M,p_y \rbrace -\frac1{4}\alpha y^2,
\]
The algebra relations are given by
\begin{gather}
\label{E5alg} [L_1,X] =-\frac{\alpha}{2},\qquad   [L_2,X]=L_1,\qquad
[L_1,L_2] =2X^3-HX, \\
\label{E5algf} X^4 -HX^2+L_1^2+\alpha L_2=0.
\end{gather}
The algebra relations (\ref{E5alg}), (\ref{E5algf}) are realized by $H=E$ and the following operators,
\begin{gather*} X =t, \qquad L_1=-\frac{\alpha}{2}\partial_t,\qquad
L_2 =-\frac1{4}\alpha\partial_t^2+\frac1{\alpha}t^2\big(E-t^2\big).
\end{gather*}
Here the only nontrivial eigenfunction is that of $L_2$ which gives solutions of the triconf\/luent Heun equation.
This model gives no obvious f\/inite dimensional restrictions.

\subsection{E4}
\label{E4}

The Hamiltonian is
\[ H =\Delta+ \alpha(x+iy).
\]
A basis for the symmetry operators is formed by $H, $ $X=p_+$ and
\[
L_1=p_x^2+\alpha x, \qquad L_2=\frac12\lbrace M,p_{+}\rbrace +\frac{i\alpha}{4}(x+iy)^2,
\]
The algebra relations are given by
 \begin{gather}\label{E4alg} [L_1,X] =-\alpha,\qquad  [L_2,X]=iX^2,\qquad
 [L_1,L_2] =-iX^3-iHX+i\left \lbrace L_1,X\right \rbrace,\\
\label{E4algf}  X^4 -\frac{2}{3} \left\lbrace L_1,X,X\right\rbrace +2HX^2+H^2+4\alpha iL_2=0.
\end{gather}
The algebra relations (\ref{E4alg}), (\ref{E4algf}) are realized by $H=E$ and the following operators
\begin{gather*}
 X=i\alpha t, \qquad L_1=i\partial_t+\frac{E}{2},\qquad L_2=-\alpha t^2\partial_t+\frac{i \alpha^4 t^4-4\alpha^2t+iE^2}{4\alpha}.
 \end{gather*}
This model gives no obvious f\/inite dimensional restrictions and the basis for the representation can be chosen as exponential functions.

\section{Conclusion}\label{section5}
In this paper, we have demonstrated at least one function space realization for each class of St\"ackel equivalence superintegrable systems. We have shown that these quadratic algebras are related to various families of special functions and orthogonal polynomials.

There are several possible directions that this analysis can be extended and the model described above can be further applied and analyzed. One possible direction, is to use the fact that there exist appropriately chosen limit processes which takes superintegrable system S9 on the sphere to all other superintegrable systems in~2D  (see e.g.~\cite{KMR,KKWMPOG}). It would be interesting to see how these limits are manifested at the level of the algebras and particularly the models. Also of interest would be to consider the restrictions from nondegenerate to degenerate systems and to study the ef\/fect of such a limit on the models.

These methods can also be applied to the construction and analysis of quadratic algebras generated by superintegrable systems in higher dimensions. In particular,  the quadratic algebras for all real nondegenerate 3D superintegrable systems were shown to contain various copies of quadratic algebras from two dimensional systems~\cite{DT2007}. As was the case in~\cite{KMPost2010, KMPost3d}, the models for the 2D system can be used as a basis for the models of the 3D system. It would be interesting to see how the algebra decomposed with respect to the subalgebras and whether that has any ef\/fect on the models obtained and possible special function relations.

\subsection*{Acknowledgements}

S.P.\ acknowledges a postdoctoral fellowship awarded by the Laboratory of Mathematical Physics of the Centre de Recherches Math\'ematiques, Universit\'e de Montr\'eal.

\pdfbookmark[1]{References}{ref}
\LastPageEnding

\end{document}